\documentclass[manuscript]{aastex}

\newcommand{\myemail}{petri@astro.u-strasbg.fr}

\shorttitle{A model for the HF-QPOs in accreting BHs}
\shortauthors{P\'etri}

\begin{document}

\title{A new model for QPOs in accreting black holes: application to
  the microquasar GRS~1915+105}

\author{J. P\'etri}
\affil{Observatoire Astronomique de Strasbourg, 
  11, rue de l'Universit\'e, 67000 Strasbourg, France.}
\affil{Centre d'\'etude des Environnements Terrestre et
  Plan\'etaires, 10-12 avenue de l'Europe, 78140 V\'elizy, France.}
\affil{Laboratoire de Radio-Astronomie, \'Ecole Normale Sup\'erieure,
  24 rue Lhomond, 75005 Paris, France.}

\email{\myemail}

\begin{abstract}
  In this paper we extend the idea suggested previously by
  \citet{Petri2005a,Petri2005b} (paper~I and II) that the high
  frequency quasi-periodic oscillations (HF-QPOs) observed in low-mass
  X-ray binaries (LMXBs) may be explained as a resonant oscillation of
  the accretion disk with a rotating asymmetric background
  (gravitational or magnetic) field imposed by the compact object.
  Here, we apply this general idea to black hole binaries.  It is
  assumed that a test particle experiences a similar parametric
  resonance mechanism such as the one described in paper~I and II but
  now the resonance is induced by the interaction between a spiral
  density wave in the accretion disk, excited close to the innermost
  stable circular orbit, and vertical epicyclic oscillations. We use
  the Kerr spacetime geometry to deduce the characteristic frequencies
  of this test particle. The response of the test particle is maximal
  when the frequency ratio of the two strongest resonances is equal to
  3:2 as observed in black hole candidates. Finally, applying our
  model to the microquasar GRS~1915+105, we reproduce the correct
  value of several HF-QPOs.  Indeed the presence of the
  168/113/56/42/28~Hz features in the power spectrum time analysis is
  predicted.  Moreover, based only on the two HF-QPO frequencies, our
  model is able to constrain the mass~$M_{\rm BH}$ and angular
  momentum~$a_{\rm BH}$ of the accreting black hole. We show the
  relation between $M_{\rm BH}$ and $a_{\rm BH}$ for several black
  hole binaries. For instance, assuming a black hole weakly or mildly
  rotating, i.e.  $a_{\rm BH} \le 0.5 \, G\,M_{\rm BH}/c^2$, we find
  that for GRS~1915+105 its mass satisfies $13 \, M_\odot \le M_{\rm
    BH} \le 20 \,M_\odot$.  The same model applied to two other
  well-known BHCs gives for GRO~J1655-40 a mass $5 \, M_\odot \le
  M_{\rm BH} \le 7 \, M_\odot$ and for XTE~J1550-564 a mass $8 \,
  M_\odot \le M_{\rm BH} \le 11 \, M_\odot$. This is consistent with
  other independent estimations of the black hole mass.  Finally for
  H1743-322, we found the following bounds, $9 \, M_\odot \le M_{\rm
    BH} \le 13 \, M_\odot$.
\end{abstract}


\section{Introduction}

High frequency quasi-periodic oscillations (HF-QPOs) are common
features to all accreting compact objects, be they neutron stars,
white dwarfs or black holes. A number of recent observations have
revealed the existence of these HF-QPOs in several black hole binaries
\citep{Strohmayer2001a,MacClintock2003,Remillard2006}. Usually, a pair
of HF-QPOs appears in a 3:2 ratio. If these oscillations are connected
to the orbital motion of the accretion disk at its inner edge as
predicted by several models, these QPOs become a useful test of
gravity in the strong field regime.

The 3:2 ratio was first noticed by \cite{Abramowicz2001}. In order to
explain this ratio, they introduced a resonance mechanism between
orbital and epicyclic motion around Kerr black holes therefore leading
to an estimate of their mass and spin. \cite{Kluzniak2004a} showed
that the twin kHz-QPOs is explained by a non linear resonance in the
epicyclic motion of the accretion disk.  \cite{Rebusco2004} developed
the analytical treatment of these oscillations.  This parametric
epicyclic resonance model of hydrodynamical modes in the accretion
disk was applied by \cite{Kluzniak2002} to some microquasars. They
pointed out the 3:2 ratio in the HF-QPOs observed in GRO J1655-40, XTE
J1550-564 and GRS1915+105. Moreover, in this latter black hole binary,
\cite{Strohmayer2001b} reported another pair of frequencies, namely
69.2~Hz and 41.5~Hz, which are in a 5:3 ratio as noticed by
\cite{Kluzniak2002} and supports the parametric resonance model.
Furthermore, \cite{Rezzolla2003} suggested that the HF-QPOs in black
hole binaries are related to p-mode oscillation in a non Keplerian
torus.

Nevertheless, the propagation of the emitted photons in curved
spacetime can also produce some intrinsic peaks in the Fourier
spectrum of the light curves \citep{Schnittman2004}. In combination
with a vertically oscillating torus, the gravitational lensing effect
can also reproduce the 3:2 ratio \citep{Bursa2004,Schnittman2005}.

Resonances in the geodesic motion of a single particle have been
investigated by \cite{Abramowicz2003}. The specific coupling force
between radial and vertical oscillation was left unspecified. Their
results for non-linear resonance were applied to accreting neutron
stars.

In this paper, we describe a coupling between spiral density waves in
the accretion disk and epicyclic motions of test particles. It is
divided in two parts. In Sec.~\ref{sec:Model} we specify the
perturbation pattern used in our model. Then the equation of motions
due to the perturbation are derived leading to some resonance
conditions.  In Sec.~\ref{sec:Discussion}, the results are applied to
GRS~1915+105 for which several components in the Fourier time analysis
are predicted in agreement with the HF-QPOs detected for this object.
We also put some constrain on the mass-spin relation for several BHCs.

\section{THE MODEL}
\label{sec:Model}

In this section, we describe the main features of the model, starting
with a simple treatment of the accretion disk, assumed to be made of
non interacting single particles orbiting in the equatorial plane of
the black hole as was already done in the previous studies (see
paper~I and II). We again neglect the hydrodynamical aspects of the
disk such as pressure. However, a detailed (magneto-)hydrodynamical
treatment of the response of the disk to gravitational or magnetic
perturbations has been given in \citet{Petri2005c,Petri2006}. For a
first approach to the problem, we neglect this refinement. Particles
evolve in the unperturbed stationary background gravitational field of
the black hole. In this simplistic approach, we want to point out the
effect of a spiral density wave propagating radially outwards in the
accretion disk. In order to account for general-relativistic effects,
we use the Kerr metric to perform our calculations. We will show that
a local outgoing one-armed m-mode will induce a beat with the test
particles and excite vertical oscillations parametrically.  Indeed,
\citet{Mao2008} showed that oscillatory modes are excited close to the
innermost stable circular orbit (ISCO) and propagate outwards to
several tens of Schwarzschild radii with constant frequency (nearly
equal to the maximum radial epicyclic frequency) and without
attenuation. They also found that these oscillations are locally
super-Keplerian.  Nevertheless, they studied only axisymmetric modes
in a pseudo-Newtonian gravitational field. A more detailed study
including full general relativity and asymmetric modes propagation
would help to give precise quantitative results.  In this paper, we
will assume that such asymmetric modes exist and exhibit the same
properties as the modes found in \citet{Mao2008}.

\subsection{Equation of motion for a test particle}

We use the same procedure as the one described in paper~I. However, in
order to take into account the spiral structure of the density wave of
azimuthal mode number~$m$ and frequency~$\Omega_{\rm w}$, we have to
change the equation of motion Eq.~(7) in paper~I by modifying the
terms ``$\cos( m \, ( \Omega - \Omega_{\rm w} ) \, t )$'' in such a
way that the spatio-temporal dependency of the argument of ``$\cos$''
implies a propagation of a pattern of azimuthal mode~$m$ at the sound
speed~$c_{\rm s}$. We remember that $\Omega$ is the local orbital
frequency in the accretion disk.  Therefore, in cylindrical
coordinates $(r,\varphi,z)$, we choose a phase dependency like $\psi =
m \, \varphi - \Omega_{\rm w} \, ( t - r / c_{\rm s} )$, where $t$ is
the time.  For a particle in circular orbit at a frequency $\Omega$,
we have
\begin{equation}
  \label{eq:PhaseApprox}
  \psi = m \, \Omega \, t - \Omega_{\rm w} \, ( t - r / c_{\rm s} )
\end{equation}
This means that viewed from the orbiting particle frame of reference,
the perturbation is rotating at the speed~$\Omega_p = |m \, \Omega -
\Omega_{\rm w}|$.  We find again a harmonic oscillator with a periodic
variation in the eigenfrequency of the system, driven by an external
periodic force generated by the perturbation.  The modulation being
sinusoidal, the Hill equation specializes to the Mathieu equation, a
well known ordinary differential equation extensively studied in
mathematical physics \citep{Morse1953}.

\subsection{Resonance conditions}

The rotation of the asymmetric pattern induces a sinusoidal variation
of the vertical epicyclic frequency~$\kappa_{\rm z}$ leading to the
well known Mathieu equation for a given azimuthal mode $m$.  To take
into account general-relativistic effects, we use the characteristic
orbital and epicyclic frequencies for the Kerr spacetime geometry,
namely
\begin{itemize}
\item the orbital frequency~$\Omega = \frac{\sqrt{G\,M_{\rm
        BH}}}{r^{3/2} + a_{\rm BH}\,\sqrt{R_g}} = \frac{c^3}{G\,M_{\rm
      BH}} \, \frac{1}{\tilde{r}^{3/2} + \tilde{a}} $
\item the radial epicyclic frequency~$\kappa_{\rm r} = \Omega \,
  \sqrt{1 - \frac{6}{\tilde{r}} + 8 \,
    \frac{\tilde{a}}{\tilde{r}^{3/2}} - 3 \,
    \frac{\tilde{a}^2}{\tilde{r}^2} }$
\item the vertical epicyclic frequency~$\kappa_{\rm z} = \Omega \,
  \sqrt{1 - 4 \, \frac{\tilde{a}}{\tilde{r}^{3/2}} + 3 \,
    \frac{\tilde{a}^2}{\tilde{r}^2}}$
\end{itemize}
We introduced the mass of the black hole $M_{\rm BH}$, its
gravitational radius $R_g = G\,M_{\rm BH}/c^2$, the adimensionalized
radius and angular momentum, respectively $\tilde{r} = r/R_g$ and
$\tilde{a} = a_{\rm BH}/R_g$.  We expect a parametric resonance
related to this time-varying vertical epicyclic frequency.  The
resonance condition derived from the vertical equation of motion is
\begin{equation}
  \label{eq:ResPara}
  |m \, \Omega - \Omega_{\rm w} | = 2 \, \frac{\kappa_{\rm z}}{n}
\end{equation}
where $n\ge 1$ is a natural integer.  Note that the speed of
propagation~$c_{\rm s}$ in the term ``$(t - r / c_{\rm s})$'' drops
out from the above resonance condition because it is a local
description at {\it fixed} radius $r$.

In the non-rotating black hole case, $a=0$, we have $\kappa_{\rm z} =
\Omega$. Therefore the parametric resonance conditions
Eq.~(\ref{eq:ResPara}) splits into two cases, depending on the
absolute sign
\begin{equation}
  \label{eq:ResPara2}
  \Omega = \frac{n}{m\,n\pm2} \, \Omega_{\rm w}
\end{equation}
It differs by a factor $1/m$ from the resonance condition derived in
paper~I and II. This discrepancy comes from the choice of the phase
dependency here being $\psi$ and describing a radially outward
propagating pattern. There are some other resonance mechanisms
happening in the accretion disk which are not of interest in the
present study. Nevertheless we remind them for completeness. First,
forced oscillations occur whenever the driven frequency is equal to
the free oscillation frequency
\begin{equation}
  \label{eq:ResForcee}
  |m \, \Omega - \Omega_{\rm w}| = \kappa_{\rm z}
\end{equation}
They are a special case of the above mentioned parametric resonance,
corresponding to $n=2$.  Second, for the corotation resonance to
happen, we should have
\begin{equation}
  \label{eq:ResCorot}
  \Omega = \Omega_{\rm w}
\end{equation}
As a consequence, we find again the same resonance mechanisms as in
paper~I and II. Eq.~(\ref{eq:ResPara2}) is also a good approximation
for slowly rotating black holes ($\tilde{a}\ll1$) because in this case
$\kappa_{\rm z} \approx \Omega$.

We now discuss in detail the case $\tilde{a}=0$ because it is
analytically tractable and gives good insight on the resonance
frequency even in the fast rotating case. Moreover, we checked that in
case of a maximally rotating Kerr black hole, $\tilde{a}=1$, the
frequencies remains nearly the same as in the case $\tilde{a}=0$
because the resonances occur in regions far away from the ISCO, such
that in these regions of the accretion disk, $\kappa_{\rm z} \approx
\Omega$ whatever $\tilde{a}$.

The predicted QPO frequency ratio $\Omega / \Omega_{\rm w}$ is shown
in Table~\ref{tab:Table} for the first three azimuthal
numbers~$m=1,2,3$ and the first four integers~$n$.  The highest
orbital frequencies where resonance occurs correspond to $m=1$ and for
the minus sign in Table~\ref{tab:Table}.  They are equal to
$3\,\Omega_{\rm w}$ and $2\,\Omega_{\rm w}$. Therefore the ratio of
the strongest oscillations are in the ratio 3:2 as observed in several
black hole binaries. This particular ratio is not a direct consequence
of the motion in general relativity but rather an intrinsic property
of the parametric resonance. General relativity is only needed in
order to describe this resonance mechanism correctly in the strong
gravity regime.  The third strongest resonance occurs when
$\Omega=\Omega_{\rm w}$ overlapping the corotation resonance.
However, if the low azimuthal numbers possess the strongest amplitude
in the perturbation power spectrum, the $m=2$ and $m=3$ resonances are
expected to be weaker. The most significant QPO frequencies are
therefore ordered in the series 3:2:1, the ``1'' only appearing as a
weak feature compared to the 3:2 HF-QPOs pair.  A quantitative
application of the model follows in the next section.

\section{DISCUSSION}
\label{sec:Discussion}

\subsection{Application to GRS~1915+105}
\label{sec:Application}

We test our model with the well studied black hole GRS~1915+105 for
which numerous observations of QPO features are available.  Several
types of QPOs can be identified in this binary system; low frequency
QPOs (LF-QPOs) ranging roughly from 1 to 10~Hz, HF-QPOs for
frequencies larger than roughly 70~Hz and very low frequency QPOs with
frequencies less than 1~Hz \citep{Fender2004}.  HF-QPO frequencies are
detected at $\nu_1 = 113$~Hz and $\nu_2 = 165$~Hz
\citep{Remillard2006}.  Other LF-QPOs have also been detected for this
object. For instance, \citet{Strohmayer2001b} reported a 40~Hz QPO
seen sometimes simultaneously with the more stable 67~Hz QPO
\citep{Morgan1997}. QPOs have been observed with the following
increasing frequencies 27-40-56-67~Hz \citep{Belloni2001,
  vanderKlis2004}. Some LF-QPOs varying between 1 and 15~Hz have been
investigated intensively by \cite{Markwardt1999}. Their properties
seem to be different from the HF-QPOs, probably implying a different
physical mechanism at work. We emphasize that it is not the scope of
this paper to explain all the observed QPOs but just to explain or
predict the HF-QPOs. LF-QPOs should be related to Lense-Thirring
precession or to some other (magneto-)hydrodynamical modes in the
accretion disk.

Nevertheless, from the frequencies of the twin HF-QPOs, the constant
angular pattern speed of the density wave is derived from our model by
$\nu_{\rm w} = \Omega_{\rm w}/2\pi = \nu_1/2 = \nu_2/3 \approx 56$~Hz.
Putting this value of the gravitational field pattern speed into
Table~\ref{tab:Table} we get the results shown in
Table~\ref{tab:TableRes}.  There is a problem on selecting the
relevant and meaningful frequencies in this table.  Indeed, the spiral
density wave travels outwards to several tens of Schwarzschild radii.
According to~\citet{Mao2008}, the precise extension of the propagation
of the wave with constant frequency and without attenuation depends on
the accretion rate and on the viscosity (at least for their
axisymmetric hydrodynamical modes).  Assuming that the wave can travel
only up to a radius $r_{\rm out}$, for a black hole of mass $M_{\rm
  BH}$ and angular momentum $a_{\rm BH}$, the lowest excited frequency
is
\begin{equation}
  \label{eq:FreqMin}
  \nu_{\rm low} = \frac{\Omega_{\rm low}}{2\,\pi} = 
  \frac{32310~{\rm Hz}}{M_{\rm BH}/M_\odot} \, \frac{1}{a_{\rm BH}/R_{\rm g} + 
    (r_{\rm out}/R_{\rm g})^{3/2}}
\end{equation}
To give some orders of magnitude, let's use $M_{\rm BH} = 10 \,
M_\odot, a_{\rm BH} = 0, r_{\rm out} = 30 \, R_{\rm g}$, we find
$\nu_{\rm low} = 19.7$~Hz. Which frequencies to select depends
strongly on the spiral wave properties, how far they can propagate,
which azimuthal numbers~$m$ are excited, at which amplitude and so on.
The detailed investigation is left for future work. We can only claim
that the observed QPOs are retrieved by our model. How to select them
remains quantitatively unclear.

Let's discuss however the predicted QPO frequencies.  The first two
strongest resonances for each azimuthal mode $m$ with the plus sign
used in Eq.~(\ref{eq:ResPara2}) match with good accuracy the
frequencies observed in GRS~1915+105.  We find namely, in decreasing
order, 168.0, 112.0, 56.0, 42.0 and 28.0~Hz for the most significant
values.  Some other resonances should also appear, but due to their
higher order (higher integer $n$) and location farther away from the
inner edge of the disk (thus an amplitude of the excitation decreasing
with radius), they do not possess a significant growth rate relevant
for the study presented here.  Moreover, X-ray emission from the outer
part of the disk is fainter, thus more difficult to detect.  As a
consequence, the parametric resonance induced by a spiral density wave
passing through an accretion disk predicts the HF-QPOs as well as some
LF-QPOs (as a byproduct) in a single unified picture.  However, there
is no way to predict the angular momentum of the black hole without
some knowledge of its mass. We can derived a relation mass-spin to
constrain the BH parameters, see next section.  Nevertheless, it is
worthwhile noting that \cite{Kato2004} was lead to the conclusion that
this microquasar is well described by a non rotating geometry
indicating that the black hole angular momentum is weak, $a_{\rm BH}
\ll R_g$ by fitting the HF-QPOs and several LF-QPOs with a resonant
interaction between non linear oscillations and warp modes in the
accretion disk.  He found $a_{\rm BH} = 0-0.15\,R_{\rm g}$.

Note also that the LF-QPOs (27-41-56~Hz) are more difficult to detect
than the HF-QPOs (113-165~Hz) because they are located in regions of
the accretion disk where emission is fainter. They correspond also to
higher azimuthal modes ($m=2,3$ compared to $m=1$). Assuming that the
main wave possesses an $m=1$ structure, the amplitude of the LF-QPOs
is expected to be small compared to those of the HF-QPOs.
Furthermore, it seems that the presence of these features depend on
the emission state of the black hole. We would then expect the
accretion rate to have an influence on the efficiency of the
parametric resonance occurring in the system.  However, the constancy
of the QPOs frequency strongly supports the fact the they are closely
related to the black hole properties and independent of the flow in
the surrounding accretion disk.  A hydrodynamical general relativistic
description is required to verify this assessment and to check the
behaviour of the accretion rate on the presence of certain QPOs.  This
would be the continuation of the work begun by \citet{Mao2008}.

\subsection{Mass-spin relation for BHCs}
\label{sec:Masses}

Strictly speaking, at this stage, our model cannot estimate
independently the angular momentum and the mass of the black hole.
Nevertheless, applying it to some BHCs, we are able to give some
constrain on their mass, assuming a non-rotating or a maximally
rotating black hole.

Following the work done by~\citet{Mao2008}, we assume that the spiral
density wave is launched close to the ISCO, at the location where the
radial epicyclic frequency is maximal, note this radius $r_{\rm max}$.
The frequency of the spiral density wave is equal to the local orbital
frequency at $r_{\rm max}$, thus the speed of the density perturbation
pattern is $\Omega(r_{\rm max}, a_{\rm BH})$.  Knowing the fundamental
frequency $\Omega_{\rm w}$ in the accretion disk for a given black
hole, the resonance condition Eq.~(\ref{eq:ResPara}) puts constrains
on the relation between mass and angular momentum by imposing
$\Omega(r_{\rm max}, a_{\rm BH}) = \Omega_{\rm w}$.

\begin{figure}
  \centering
  \includegraphics{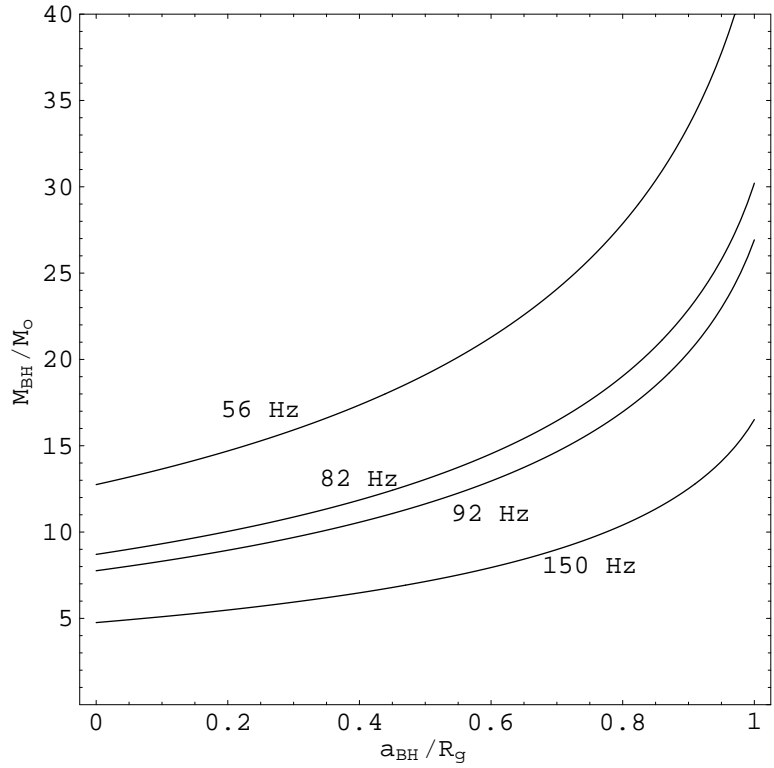}
  \caption{The relation between mass~$M_{\rm BH}$ (normalized to the
    mass of the Sun~$M_\odot$) and angular momentum $a_{\rm BH}$ for
    different fundamental excitation frequencies corresponding to the
    four BHCs, namely, GRS1915+105 with 56~Hz, GRO~J1655-40 with
    150~Hz, XTE~J1550-564 with 92~Hz and H1743-322 with 82~Hz.}
  \label{fig:MasseMomCin}
\end{figure}

This mass-angular momentum relation is shown in
Fig.~\ref{fig:MasseMomCin} for four BHCs (GRS1915+105, GRO~J1655-40,
XTE~J1550-564, H1743-322) for which the fundamental frequency is
known.  If we assume a black hole weakly or mildly rotating, i.e.
$a_{\rm BH} \le 0.5 \, G \, M_{\rm BH} / c^2 $, we find that for
GRS~1915+105 its mass satisfies $13 \, M_\odot \le M_{\rm BH} \le 20
\,M_\odot$.  The same model applied to other BHCs gives for
GRO~J1655-40 a mass $5 \, M_\odot \le M_{\rm BH} \le 7 \, M_\odot$ and
for XTE~J1550-564 a mass $8 \, M_\odot \le M_{\rm BH} \le 11 \,
M_\odot$. This is consistent with other independent estimations of the
BH mass.  Finally for H1743-322, we found the following bounds, $9 \,
M_\odot \le M_{\rm BH} \le 13 \, M_\odot$.

\section{CONCLUSION}
\label{sec:Conclusion}

In this paper, the consequences of an outgoing spiral density wave on
the evolution of a test particle orbiting around a Kerr black hole
have been explored.  The twin peaks ratio around 3:2 for the kHz-QPOs
is naturally explained by the parametric resonance model. The
connection to lower frequency QPOs is also clearly demonstrated. It is
an extension to accreting black hole for the model already suggested
in neutron star and white dwarf binaries.

From the analysis of HF-QPOs, we are also able to constrain the mass
and the spin of the hole as already suggested by \cite{Torok2005}.  In
the case of GRS~1915+105, it appears that the 56~Hz feature in the
Fourier time analysis is a fundamental frequency of the black hole
from which the other QPOs can be derived. This feature should
therefore be related to the intrinsic properties of the black, namely,
its mass and its angular momentum. The way to associate $\Omega_{\rm
  w}$ with $a_{\rm BH}$ and $M_{\rm BH}$ remains unclear and needs
further investigation but spiral density wave excitation close to the
ISCO at nearly the maximum of the radial epicyclic frequency as
suggested by~\citet{Mao2008} is a interesting idea that needs further
investigations.

\begin{acknowledgements}
  This work was partly supported by a grant from the french ANR MAGNET.  
\end{acknowledgements}

\clearpage

\begin{table}
  \begin{center}
    \caption{Ratio of orbital frequency to the perturbation angular
      pattern speed at the radii where the parametric resonance is
      expected to occur.  Left part of the table corresponds to the
      minus sign whereas the right part corresponds to the plus sign
      taken in Eq.~(\ref{eq:ResPara2}).}
    \begin{tabular}{c|cccc|cccc}
      \hline
      \hline
      \multicolumn{9}{c}{$\Omega / \Omega_{\rm w}$} \\
      \hline
      \hline
      m & \multicolumn{4}{c|}{n ($-$)} & \multicolumn{4}{c}{n ($+$)} \\
      \hline
      & 1 & 2 & 3 & 4  & 1 & 2 & 3 & 4 \\
      \hline
      1 & -1   & --- &   3  &   2 & 1/3 & 1/2 & 3/5  & 2/3 \\
      2 & ---  & 1   &  3/4 & 2/3 & 1/4 & 1/3 & 3/8  & 2/5 \\
      3 & 1    & 1/2 &  3/7 & 2/5 & 1/5 & 1/4 & 3/11 & 2/7 \\
      \hline
    \end{tabular}  
    \label{tab:Table}
  \end{center}
\end{table}

\clearpage

\begin{table}[htbp]
  \begin{center}
    \caption{QPO frequencies predicted for GRS~1915+105 
      from the orbital motion where parametric resonance occurs. 
      The most interesting frequencies are marked in bold face.}
    \begin{tabular}{c|cccc|cccc}
      \hline
      \hline
      \multicolumn{9}{c}{$\nu_\mathrm{QPO}$ (in Hz) for GRS~1915+105} \\
      \hline
      \hline
      m & \multicolumn{4}{c|}{n ($-$)} & \multicolumn{4}{c}{n ($+$)} \\
      \hline
      & 1 & 2 & 3 & 4  & 1 & 2 & 3 & 4 \\
      \hline
      1 & -56.0 & --- & {\bf 168.0} & {\bf 112.0} & 18.7 & 28.0 & 33.6 & 37.3 \\
      2 & ---   & {\bf 56.0}  & {\bf 42.0} & 37.3 & 14.0 & 18.7 & 21.0 & 22.4  \\
      3 & {\bf 56.0}  & {\bf 28.0} & 24.0 & 22.4  & 11.2 & 14.0 & 15.3 & 16.0 \\
      \hline
    \end{tabular}  
    \label{tab:TableRes}
  \end{center}
\end{table}

\end{document}